\documentclass[prd,showpacs,amsmath,amssymb,twocolumn,preprintnumbers,nofootinbib]{revtex4}

\usepackage[utf8]{inputenc}
\usepackage{graphicx}
\usepackage{amsmath}
\usepackage{hyperref}

\normalbaselines
\newcommand{\sgn}{\mathop{\rm sgn}\nolimits}

\newcommand{\erf}{\mathop{\rm erf}\nolimits}

\newcommand{\Fst}{{\mathop {\rule{0pt}{0pt}{F}}\limits^{\,*}}\rule{0pt}{0pt}}

\begin{document}

\title{Einstein-Maxwell-axion theory: Dyon solution with regular electric field}

\author{Alexander B. Balakin}
\email{Alexander.Balakin@kpfu.ru} \affiliation{Department of
General Relativity and Gravitation, Institute of Physics, Kazan
Federal University, Kremlevskaya str. 18, Kazan 420008, Russia}

\author{Alexei E. Zayats}
\email{Alexei.Zayats@kpfu.ru} \affiliation{Department of
General Relativity and Gravitation, Institute of Physics, Kazan
Federal University, Kremlevskaya street 18, Kazan 420008, Russia}

\begin{abstract}
In the framework of Einstein-Maxwell-axion theory we consider static spherically symmetric solutions, which describe a magnetic monopole in the axionic environment. These solutions are interpreted as the solutions for an axionic dyon, the electric charge of which is composite, i.e., in addition to the standard central electric charge, it includes  an effective electric charge induced by the axion-photon coupling. We focus on the analysis of that solutions, which are characterized by the electric field regular at the center. Special attention is paid to the solutions with the electric field, which is vanishing at the center, has the Coulombian asymptote and thus display an extremum at some distant sphere.  Constraints on the electric and effective scalar charges of such an object are discussed.
\end{abstract}


\maketitle

\section{Introduction}

In 1987 Wilczek has formulated the idea that for a distant observer the magnetic monopole in an axionic environment looks like a dyon with magnetic and effective electric charge \cite{Wilczek2}. This idea was based on the prediction of the axion electrodynamics that the interaction between the radial magnetic field, attributed to the monopole, and the surrounding pseudoscalar (axion) field produces the radial electric field without real electric charge at the center. That is why one can say, that Wilczek presented in 1987 the first example of the so-called axionic dyon. The axion electrodynamics, on which this result was based, has been established and developed in the decade 1977-1987, being inspired by the theoretical discovery of Peccei and Quinn  of the CP-invariance conservation \cite{PQ}, and by discussions about a new light pseudo-Goldstone boson introduced by Weinberg \cite{Weinberg} and Wilczek \cite{Wilczek}. The model of coupling of the pseudoscalar and electromagnetic fields was formulated in covariant form  by Ni in \cite{Ni77}; the axion electrodynamics written in the 3-dimensional form was used by many authors (see, e.g., the work of Sikivie \cite{Sikivie83}). Since the axions are considered to be candidates to the dark matter particles \cite{ADM1,ADM2,ADM3,ADM4,ADM5,ADM6,ADM7,ADM8,ADM9}, the physics of axions had become one of the key elements of numerous applications to cosmology and astrophysics. These applications involve into consideration various models of interaction of gravitational, electromagnetic, scalar and pseudoscalar fields, which are called nowadays the Einstein-Maxwell-axion, and Einstein-Maxwell-axion-dilaton models (see, e.g., \cite{EMAD1,EMAD2,EMAD3}). Also, these applications attract the attention to the models, which belong to the class of theories indicated by term Extended Axion Electrodynamics \cite{ExtendedAE1,ExtendedAE2,ExtendedAE3,ExtendedAE4,ExtendedAE5,ExtendedAE6,ExtendedAE7,ExtendedAE8}.

In 1991 Lee and Weinberg \cite{LW1991} studied spherically symmetric solutions for static black holes with massless axionlike scalar field; in fact, it was a realization of the Wilczek idea in the framework of Einstein-Maxwell-axion theory. Lee and Weinberg have obtained self-consistent master equations for the axion field and metric coefficients, analyzed the asymptotic properties of the solutions, and studied analytic and numeric solutions for the cases of large and small values of the constant of the axion-photon coupling. If we omit the initial electric charge at the center of the object described in \cite{LW1991}, we find the solution for the axionic dyon, which was obtained in the framework of the Einstein-Maxwell-axion model, and was predicted in \cite{Wilczek2} using the simple Maxwell-axion model. In this sense, one can say, that in \cite{Wilczek2,LW1991} the authors presented the first (static) example of the so-called Longitudinal Magneto-Electric Cluster, in which the magnetic and axionically induced electric fields are parallel to one another. Later the solutions describing the Longitudinal Clusters were found in the systems with the pp-wave symmetry \cite{pp1}, and in the context of search for fingerprints of relic axions in the terrestrial magnetosphere \cite{BG}.

Now we are interested to find a regular solution for the axionic dyon. What does this means? In 1968 Bardeen \cite{Bardeen} attracted the attention to the solutions of the field equations regular in the center. The first idea was to modify the equations for the electric field so that it will be finite at $r=0$; for instance, it might be the function $E(r)=\frac{Q}{r^2+a^2}$ with $E(0)=\frac{Q}{a^2}$ and the Coulombian asymptote $E(r)\to \frac{Q}{r^2}$. In the framework of Einstein-Maxwell theory the regularity in the center assumes that not only the electric field is finite, but the metric coefficients and all curvature invariants are finite as well. The story of search for regular solutions is worthy to be a subject of special review; we would like to mention only three details in this context. First, the nonminimal coupling of electromagnetic and gauge fields can provide the gravitational field to be regular (see, e.g., \cite{Reg1,Reg2,Reg3}). Second, the nonminimal coupling can provide the electric field to be finite in the center (see, e.g., \cite{Reg4,Reg5,Reg6}). Third, for solutions with a magnetic monopole field the situation is not perfect; for the mentioned solutions the first invariant of the electromagnetic (or gauge) field, ${\vec{B}}^2-{\vec{E}}^2$, is not regular in the center, since the magnetic field, $B(r)=\frac{\nu}{r^2}$, in contrast to the electric one, cannot be finite there. As for the second (pseudo)invariant $(\vec{B} \cdot \vec{E})$, it is possible to be finite in the center, when the electric field is not only finite, but tends to zero not slowly than $r^2$. On the other hand, if the electric field strength $E$ is finite at the origin, but does not vanish, vector field $\vec{E}$ has a hedgehog-like singularity. Therefore if we expect to find a solution, which is characterized by the electric field regular in the center in the strict sense of the word, we should require the condition $E(0)=0$ to be satisfied.

Thus, searching for the regular axionic dyons, we are faced with the problem to find an exact solution of the field equations, for which the electric component vanishes both at $r\to 0$ and $r \to \infty$.
Below we intend to show that it is possible for magnetic monopole surrounded by the pseudoscalar (axion) field, when the guiding parameters of the model are specifically coupled.

The paper is organized as follows. In Section~\ref{EMaformalism}, we remind basic details of the Einstein-Maxwell-axion theory, and recover well-known solutions with vanishing constant of the axion-photon coupling ($\gamma=0$), using the harmonic spacetime coordinates.
In Section~\ref{sec3}, we analyze the solutions with nonvanishing $\gamma$; in Subsection~\ref{exactS} we discuss an example of exact solution for the axionic dyon singular at the center; in Subsection~\ref{backgroundsolution} we study (analytically) the regular solutions of the axion electrostatics in the background of magnetic monopole; the results of the numerical study are presented in Subsection~\ref{numsolution}.
Section~\ref{conclusion} contains conclusions.

\section{Einstein-Maxwell-axion model}\label{EMaformalism}

\subsection{Basic formalism}

The action functional of the Einstein-Maxwell-axion model  takes the form
\begin{align}
    {}&S_{EMa}=\int {\cal L}\, \sqrt{-g}\, d^4x,\nonumber \\
  {}&{\cal L}=\frac{R}{2\kappa}+\frac14 F_{ik}F^{ik} +\frac{1}{4}\gamma F_{ik}\Fst^{ik}\phi\nonumber\\
  {}&\qquad{}-\frac12 \nabla_i\phi\nabla^i\phi+\frac12 m^2_a\phi^2.\label{LagrEMa}
\end{align}
Here $R$ is the Ricci scalar; $g$ is the determinant of the metric tensor $g_{ik}$; $\kappa$ is the Einstein constant, $F_{ik}$ is the Maxwell tensor, $\Fst_{ik}$ denotes its dual tensor,
$\phi$ stands for the pseudoscalar (axion) field; $\gamma$ is the constant of the axion-photon coupling; and $m_a$ is the axion mass.

The variation of the action (\ref{LagrEMa}) with respect to potentials of the electromagnetic field $A_i$, to the axion field $\phi$, to the spacetime metric $g^{ik}$ gives, respectively,
the equations of axion electrodynamics
\begin{equation}
  \nabla_iF^{ik}+\gamma \Fst^{ik}\nabla_i\phi=0,\label{MEq}
\end{equation}
the equation for the axion field $\phi$
\begin{equation}
  \nabla_i\nabla^i \phi+m^2_a\phi+\frac{\gamma}{4}F_{ik}\Fst^{ik}=0, \label{axEq}
\end{equation}
and the equations for the gravitational field
\begin{gather}
  R_{ik}-\frac12 R g_{ik}=\kappa\left(T_{ik}^{(M)}
  +T_{ik}^{(a)}\right).\label{EinEq}
\end{gather}
Here $T_{ik}^{(M)}$ and $T_{ik}^{(a)}$ are the energy-momentum tensors for the electromagnetic and axion fields, respectively, which are defined as follows
\begin{gather}
  T_{ik}^{(M)}=\frac{1}{4}g_{ik}F_{mn}F^{mn}-F_{in}{F_{k}}^n, \label{TM}\\
  T_{ik}^{(a)}=\nabla_i\phi\nabla_k\phi-\frac12 g_{ik} \nabla_n\phi\nabla^n\phi+\frac12 m_a^2\phi^2 g_{ik}.\label{Tax}
\end{gather}
The dual Maxwell tensor satisfies the equation $\nabla_k \Fst^{ik}=0$, which is free of information about the axion field.

\subsection{Static spherically symmetric spacetime}

Let us consider a static spherically symmetric spacetime with the metric
\begin{equation}
  ds^2={\rm e}^{-2\beta(u)}dt^2-{\rm e}^{2\beta(u)}\left[{\rm e}^{-4\rho(u)}du^2+{\rm e}^{-2\rho(u)}d\Omega^2\right].
  \label{metric}
\end{equation}
 We use the harmonic coordinate system \cite{Bron73}, in which the variable $u$ plays the role of a radial coordinate; the spatial infinity corresponds to $u=0$.
 We assume that the axion field depends on the radial coordinate only, i.e., $\phi=\phi(u)$.
This system is more convenient to analyze scalar field models, but when we will need to revive the usual spherical coordinate notation, we put
\begin{equation}\label{r}
  r={\rm e}^{\beta(u)-\rho(u)}.
\end{equation}
At the spatial infinity, i.e, at $u=0$, asymptotic behavior of the spacetime metric is supposed to be Minkowskian. It means that
\begin{equation}
\beta(0)=0, \quad \rho(u)|_{u\to 0}\sim \ln u. \label{ic1}
\end{equation}
In this paper, we focus on the study of configurations with a magnetic monopole located at the center; the Maxwell tensor components are chosen to be equal to
\begin{equation}
  F_{ut}=q(u){\rm e}^{-2\beta(u)},\quad F_{\theta\varphi}=\mu \sin\theta,\label{M}
\end{equation}
where the constant $\mu$ relates to the magnetic charge, $q(u)$ is a function to be found. To characterize the electric field it is convenient to introduce also the scalar quantity $E$ defined as follows
\begin{equation}\label{Edef}
  E\equiv \frac{q}{r^2}=q\,{\rm e}^{2\rho-2\beta}.
\end{equation}
In fact, the scalar $E$ is the tetrad component of the Maxwell tensor $E=\sqrt{-F_{ut}F^{ut}}$.
In  these terms the equations of axion electrodynamics (\ref{MEq}) reduce to one equation
\begin{gather}
\left(q-\gamma\mu\phi\right)'_u=0,
\end{gather}
yielding the solution
\begin{gather}
q=Q+\gamma\mu\left(\phi-\phi_0\right).\label{Eqaq}
\end{gather}
The constant of integration  $\phi_0$ is the value of the axion field at the infinity, i.e., $\phi(u=0)=\phi_0$; similarly, we define $Q=q(0)$.

The axion field equation (\ref{axEq}) takes now the form
\begin{equation}
\phi''_{uu}-\gamma\mu q\,{\rm e}^{-2\beta}-m^2_a \phi\, {\rm e}^{2\beta-4\rho} =0.\label{Eqa1}
\end{equation}
Using (\ref{Eqaq}) we can rewrite this equation as follows:
\begin{equation}
q''_{uu}-\gamma^2\mu^2 q\,{\rm e}^{-2\beta}-m^2_a \left(q+\gamma\mu\phi_0-Q\right)\, {\rm e}^{2\beta-4\rho}=0.\label{Eqq}
\end{equation}

There are four nontrivial equations of the gravitational field. For the metric (\ref{metric}) four nonvanishing components of the Einstein tensor $G^k_i = R^k_i -\frac12 \delta_i^k R$ are
\begin{gather}
G_u^u={\rm e}^{-2\beta+4\rho}\left({\beta'_u}^2-{\rho'_u}^2+{\rm e}^{-2\rho}\right),\\
G_\theta^\theta=G_\varphi^\varphi={\rm e}^{-2\beta+4\rho}\left(-{\beta'_u}^2+{\rho'_u}^2+\rho''_{uu}\right),\\
G_t^t={\rm e}^{-2\beta+4\rho}\left(-{\beta'_u}^2+{\rho'_u}^2-2\beta''_{uu}+2\rho''_{uu}+{\rm e}^{-2\rho}\right).
\end{gather}
The corresponding four nonvanishing components of the energy-momentum tensor $T^k_i = T^{k(M)}_i+T^{k(a)}_i$  take the form (see (\ref{TM}) and (\ref{Tax}))
\begin{gather}
T_u^u=\frac12{\rm e}^{-2\beta+4\rho}\left[(\mu^2+q^2){\rm e}^{-2\beta}-{\phi'_u}^2\right]+\frac12 m^2_a\phi^2,\\
T_\theta^\theta=T_\varphi^\varphi=\frac12{\rm e}^{{-}2\beta{+}4\rho}\left[{-}(\mu^2+q^2){\rm e}^{{-}2\beta}{+}{\phi'_u}^2\right]{+}\frac12 m^2_a\phi^2,\\
T_t^t=\frac12{\rm e}^{-2\beta+4\rho}\left[(\mu^2+q^2){\rm e}^{-2\beta}+{\phi'_u}^2\right]+\frac12 m^2_a\phi^2.
\end{gather}
If we assume (as in \cite{LW1991}) that the axion field is massless, $m_a=0$, three independent equations  for gravity field can be rewritten as
\begin{gather}
\rho''_{uu}+{\rm e}^{-2\rho}=0, \label{Eq1} \\
\beta''_{uu}+\frac{\kappa}{2}\left(\mu^2+q^2\right){\rm e}^{-2\beta}=0,\label{Eq2}\\
{\rho'_u}^2-{\rm e}^{-2\rho}={\beta'_u}^2+\frac{\kappa}{2}{\phi'_u}^2-\frac{\kappa}{2}\left(\mu^2+q^2\right){\rm e}^{-2\beta}.\label{Eq3}
\end{gather}
Clearly, the first equation is decoupled from other ones, and can be immediately resolved. Indeed, the first integral of (\ref{Eq1}) is
\begin{equation}
{\rho'_u}^2-{\rm e}^{-2\rho}=C,\label{Eq3a}\\
\end{equation}
and the the solution satisfying the condition (\ref{ic1}) takes the form
\begin{gather}
 \rho=\ln \Pi(C,u),\nonumber \\
  \Pi(C,u)\equiv\left\{\begin{array}{l} \dfrac{\sinh \nu u}{\nu },\ \hbox{when $C=\nu^2>0$},\medbreak\\
  u,\ \hbox{when $C=0$},\medbreak\\ \dfrac{\sin \nu u}{\nu},\ \hbox{when $C=-\nu^2<0$}.\end{array}\right.\label{rho}
\end{gather}
Also, one can check directly, that with (\ref{rho}) Eq.~(\ref{Eq2}) is a differential consequence of (\ref{Eq3}) with (\ref{Eqaq}).
Thus the key subsystem of master equations consists of the following pair of equations
\begin{gather}
{\beta'_u}^2+\frac{\kappa}{2\gamma^2\mu^2}{q'_u}^2-\frac{\kappa}{2}\left(\mu^2+q^2\right){\rm e}^{-2\beta}=C,\nonumber\\
q''_{uu}-\gamma^2\mu^2 q\,{\rm e}^{-2\beta}=0.
\label{key}
\end{gather}
When the quantities $\beta(u)$ and $q(u)$ are found, the axion field and the electric field can be reconstructed as
\begin{equation}
\phi=\phi_0+\frac{q-Q}{\gamma\mu} \,, \quad E(u) =  q(u) \,{\rm e}^{-2\beta} \Pi^2(C,u) \,.
\label{nm}
\end{equation}
In other words, we have to find two functions $\beta$ and $q$, which satisfy the key system of equations (\ref{key}).
Since we use nonstandard coordinate $u$ instead of radial variable $r$, we would like to comment how the known solutions can be displayed in these terms.

\subsection{Known solutions in the $u$-representation with vanishing constant of axion-photon coupling}

In order to illustrate a behavior of the metric functions $\beta$, $\rho$ and the function $q$, let us give some examples of well-known spacetimes, for which the axion-photon coupling is supposed to be absent, $\gamma=0$.

\subsubsection{Schwarzschild solution}

In case when $\mu=0$ and $q=0$, the first equation from Eq.~(\ref{key}) reduces to the following form
\begin{gather}
 {\beta'_u}^2=C>0
\end{gather}
and the solution to it with the condition (\ref{ic1}) can be found immediately
\begin{equation}
\beta=Mu,
\end{equation}
where $C=M^2$. The formula (\ref{rho}) gives
\begin{equation}
\rho=\ln\left(\frac{\sinh Mu}{M}\right).
\end{equation}
Thus, we obtain the Schwarzschild metric in the harmonic coordinates
\begin{equation}
  ds^2={\rm e}^{-2Mu}dt^2-\frac{M^4\,{\rm e}^{2Mu}}{\sinh^4 Mu}du^2-\frac{M^2\,{\rm e}^{2Mu}}{\sinh^2 Mu}d\Omega^2.
\end{equation}
After transformation of the radial coordinate (see (\ref{r}))
$$
r=\frac{M\,{\rm e}^{Mu}}{\sinh Mu}, \quad \Leftrightarrow \quad u=-\frac{1}{2M}\ln\left(1-\frac{2M}{r}\right)
$$
this metric returns to its standard form
$$
ds^2 = \left(1-\frac{2M}{r} \right)dt^2 - \left(1-\frac{2M}{r} \right)^{-1}dr^2 - r^2 d\Omega^2
\,.
$$
The constant $M$ plays here the role of the mass. Mention should be made, that the $u$-coordinate system covers the Schwarzschild spacetime from the spatial infinity ($u=0$) till the horizon ($u\to\infty$) only. When $u\to\infty$ the metric component $g_{tt}={\rm e}^{-2Mu}$ tends to zero, i.e., $r\to 2M$.

\subsubsection{Reissner-Nordstr\"om solution}

Let the axion field and the function $q$ be constant, $\phi=\phi_0$, $\gamma=0$ and $q=Q$. Then the second equation from Eq.~(\ref{key}) is an identity, and the first one is simplified as \begin{gather}
 {\beta'_u}^2-\frac{\kappa}{2}\left(\mu^2+Q^2\right){\rm e}^{-2\beta}=C.
\end{gather}
The solution to this equation, which satisfies the condition $\beta(0)=0$, is
\begin{equation}
 \beta=\ln\left[\frac{\Pi(C,u+u_*)}{\Pi(C,u_*)}\right],
\end{equation}
where the value $u_*$ can be obtained from the condition
\begin{equation}
\Pi(C,u_*)=\left[\frac{\kappa(\mu^2+Q^2)}{2}\right]^{-1/2}.
\end{equation}
In order to clarify the sense of the constant $C$ for the Reissner-Nordstr\"om solution, we consider the case $u\to0$.
At the origin the metric function $\beta$ behaves as
\begin{equation}
  \beta|_{u\to0}\sim \left[C+\frac{1}{\Pi(C,u_*)^2}\right]^{1/2}\cdot u,
\end{equation}
and keeping in mind the Schwarzschild solution we can identify the factor in front of $u$ with the mass $M$, i.e.,
\begin{equation}
  C=M^2-\frac{\kappa(\mu^2+Q^2)}{2}.
\end{equation}
Thus, the Reissner-Nordstr\"om solution in the harmonic coordinate system takes the form
\begin{align}
  ds^2&=\frac{\Pi(C,u_*)^2}{\Pi(C,u+u_*)^2}dt^2\nonumber\\
  {}&{}-\frac{\Pi(C,u+u_*)^2}{\Pi(C,u)^2\,\Pi(C,u_*)^2}\left(\frac{du^2}{\Pi(C,u)^2}+d\Omega^2\right).
  \label{RNa}
\end{align}
We have to compare this solution with the well-known one
\begin{gather}
  ds^2= A(r) dt^2 - \frac{dr^2}{A(r)}- r^2 {d\Omega}^2 \,,\nonumber\\
  A(r) = \left(1-\frac{M}{r}\right)^2 + \frac{1}{2r^2}(Q^2+\mu^2-2M^2).  \label{RN}
\end{gather}
The solutions (\ref{RNa}) can be identified with (\ref{RN}) keeping in mind the number of horizons.

\noindent
{\it (i)}  One horizon.

\noindent
When $M^2>\kappa(Q^2+\mu^2)/2$, i.e., when $C=\nu^2>0$, there is one horizon at $u=\infty$ and $r(u\to\infty)=M+\sqrt{C}$.

\noindent
{\it (ii)}  Naked singularity.

\noindent
When $C=-\nu^2<0$, one obtains that
$$
  {\rm e}^\rho=\frac{\sin \nu u}{\nu}, \quad
 {\rm e}^\beta=\frac{\sin \nu (u+u_*)}{\sin \nu u_*},
 $$
 $$
 u_*=\frac{1}{\nu}\arcsin\frac{\nu}{\sqrt{\nu^2+M^2}},
$$
\begin{equation}
  r={\rm e}^{\beta-\rho}=\sqrt{\nu^2+M^2}\cdot\frac{\sin \nu(u+u_*)}{\sin \nu u}.
  \label{dd}
\end{equation}
If $u+u_*=\pi/\nu$ then $r\to 0$, therefore this point $u=u_*$
corresponds to the central naked singularity.

\noindent
{\it (iii)}  Double horizon.
 \noindent
 When $M^2=\kappa(Q^2+\mu^2)/2$, i.e., when $C=0$, we obtain
\begin{equation}
  {\rm e}^\rho=u \,, \quad \quad
 u_*=\frac{1}{M}, \quad {\rm e}^{-2\beta}= \left(1-\frac{M}{r} \right)^2 \,, \
\end{equation}
After transformation of the radial coordinate

\begin{equation}
r=\frac{(1+Mu)}{u}\quad \Leftrightarrow \quad u=\frac{1}{r-M}
\end{equation}
one can derive the standard form of the metric
\begin{equation}
  ds^2=\left(1-\frac{M}{r}\right)^2dt^2-\left(1-\frac{M}{r}\right)^{-2}dr^2-r^2 d\Omega^2.
\end{equation}

\subsubsection{Penney and Fisher solutions}

When the axion-photon coupling constant $\gamma$ is equal to zero and $m_a=0$, the Eq.~(\ref{Eqa1}) reduces to $\phi''_{uu}=0$, thus the axion field is linear in the variable $u$
\begin{equation}\label{a}
  \phi=\phi_0+Pu \,.
\end{equation}
The integration constant $P$ can be indicated as an scalar (axion) ``charge''. The constant $C$ is now a combination of the charges $P$, $Q$, and $\mu$
\begin{equation}
  C=M^2-\frac{\kappa}{2}(\mu^2+Q^2-P^2).\label{C}
\end{equation}
The equations (\ref{Eq3}) give now
\begin{equation}\label{k}
\beta_u^2=M^2+\frac{\kappa}{2}(\mu^2+Q^2)\left[{\rm e}^{-2\beta}-1\right],
\end{equation}
and the solution to this equation takes the form
\begin{equation}\label{k1}
 \beta=\ln\frac{\Pi(\tilde{C},u+u_*)}{\Pi(\tilde{C},u_*)}.
\end{equation}
Here the modified constant $\tilde{C}$ is of the form
\begin{gather}
\tilde{C}=M^2-\frac{\kappa}{2}(\mu^2+Q^2),\\
\Pi(\tilde{C},u_*)=\left(\frac{\kappa(\mu^2+Q^2)}{2}\right)^{-1/2}.
\end{gather}
For the metric functions $\beta$ and $\rho$ given by (\ref{rho})) the linear element (\ref{metric}) covers the Penney solution \cite{Penney}
\begin{align}
  ds^2&=\frac{\Pi(\tilde{C},u_*)^2}{\Pi(\tilde{C},u+u_*)^2}dt^2\nonumber\\
  {}&{}-\frac{\Pi(\tilde{C},u+u_*)^2}{\Pi(C,u)^2\,\Pi(\tilde{C},u_*)^2}\left(\frac{du^2}{\Pi(C,u)^2}+d\Omega^2\right).\label{Pen}
\end{align}
Clearly, when $P=0$ the constants $C$ and $\tilde{C}$ coincide, and the Penney solution reduces to the Reissner-Nordstr\"om one.

When $\tilde{C}=0$, i.e.,
\begin{equation}\label{k2}
M^2=\frac{\kappa}{2}(Q^2+\mu^2) \,, \quad C=\kappa P^2/2>0 \,, \quad u_*=\frac{1}{M} \,,
\end{equation}
we recover the ``extremal'' Penney solution
\begin{equation}\label{f}
  ds^2=\frac{dt^2}{(1+Mu)^2}-\frac{C(1+Mu)^2}{\sinh^2 \sqrt{C}u}\left(\frac{C\, du^2}{\sinh^2 \sqrt{C}u}+d\Omega^2\right).
\end{equation}

In the particular case, if both electric and magnetic charges, $Q$ and $\mu$, vanish, the metric (\ref{Pen}) turns into
the Fisher metric \cite{Fisher}
\begin{equation}
  ds^2={\rm e}^{-2Mu}dt^2-\frac{{\rm e}^{2Mu}}{\Pi(C,u)^4}du^2-\frac{{\rm e}^{2Mu}}{\Pi(C,u)^2}d\Omega^2,
\end{equation}
where the constant
\begin{equation}
C=M^2-\frac{\kappa P^2}{2}
\end{equation}
can be positive, vanishing or negative depending on the relation between the mass $M$ and the scalar (axion) charge $P$.

\section{Solutions with nonvanishing constant of the axion-photon coupling, $\gamma \neq 0$ }\label{sec3}

Let us consider the general case, for which the axion-photon
coupling constant $\gamma$ does not vanish. We deal now with the key system of equations
\begin{gather}
{\beta'_u}^2+\frac{\kappa}{2\gamma^2\mu^2}{q'_u}^2-\frac{\kappa}{2}\left(\mu^2+q^2\right){\rm e}^{-2\beta}=C,\nonumber\\
q''_{uu}-\gamma^2\mu^2 q\,{\rm e}^{-2\beta}=0,\label{keyrep1}
\end{gather}
with the the boundary conditions
\begin{equation}
  \beta(0)=0, \  \beta'_u(0)=M ,  \  q(0)=Q  , \ q'_u(0)=\gamma \mu P \,.\label{k1}
\end{equation}
The first condition for $\beta$ is the requirement that the spacetime is asymptotically Minkowskian; the second one introduces the asymptotic Schwarzschild mass $M$.
The first condition for $q$ means that $Q$ is the asymptotic electric charge. As for the last condition, it appears from the relationship $\phi=\phi_0+\frac{q-Q}{\gamma\mu}$ and the definition for the axion charge $\phi'_u(0)=P$. As usual, we denote the asymptotic value of the pseudoscalar (axion) field as   $\phi(0)=\phi_0$.
and for this version of the key system of equations the constant $C$ is not arbitrary, it satisfies the condition (\ref{C})
$C=M^2-\frac{\kappa}{2}\left(\mu^2+Q^2-P^2\right)$.

Clearly, the key system of equations (\ref{keyrep1}) does not depend on $u$ explicitly, we see $u$ only as the argument of $\beta(u)$ and $q(u)$.
This means that we can search for particular solutions of the form $\beta=\beta(q(u))$, and replace the derivative $\beta^{\prime}_{u}$ by $\frac{d\beta}{dq} q^{\prime}_{u}$ in the key system yielding the following equation:
\begin{equation} \label{vv}
  \beta''_{qq}=-\frac{\frac12 \kappa(\mu^2+q^2)+\gamma^2\mu^2q\beta'_q}{C{\rm e}^{2\beta}+\frac12\kappa(\mu^2+q^2)}\ \left({\beta'_q}^2+\frac{\kappa}{2\gamma^2\mu^2}\right)\,.
\end{equation}
We will use this consequence in the next subsection to obtain particular exact solution to the key system.

\subsection{Exact solution with the singularity at the center}\label{exactS}

\begin{figure}[t]
\centerline{\includegraphics[height=7cm]{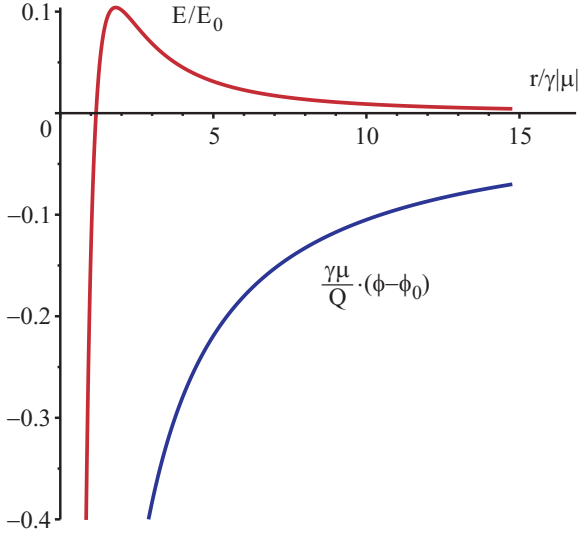}} 
\caption{Behavior of the electric and axion fields described by exact solutions (\ref{explicitsol2}), (\ref{explicitsolr}). The upper curve is the plot of the normalized electric field $\frac{E}{E_0}$, where $E_0=\frac{Q}{\gamma^2\mu^2}$, as the function of the normalized radial variable $\frac{r}{\gamma|\mu|}$. The lower curve relates to the normalized axion field $\frac{\gamma \mu}{Q}(\phi-\phi_0)$. For presented plots we have chosen $|Q|=|\mu|$.}\label{Fig1}
\end{figure}

In general case the key system of equations admits the numerical study only, that is why we would like to start our discussion with a particular but explicit example of a solution, when the constant $C$ is vanishing, $C=0$.  Then the first equation (\ref{keyrep1}) admits the solution quadratic in $q$:
\begin{equation}
\beta(q)=\frac{Q^2-q^2}{2\mu^2},
\label{explicitsol1}
\end{equation}
when five parameters $M$, $Q$, $\gamma$, $\kappa$, $\mu$ satisfy the following three relationships
\begin{equation}
M=\gamma |Q|, \quad \kappa=2\gamma^2 \,, \quad  P=-\mu\sgn Q \,.
\label{ex2}
\end{equation}
Since $C=0$, the second metric coefficient is of the form $\rho(u)=\ln u$.
In order to find the function $q(u)$ we focus on the second equation  (\ref{keyrep1}). With the parameters given by (\ref{ex2}) and boundary conditions (\ref{k1}) the first integral of that equation is
\begin{equation}
  {q'_{u}}^2 - \gamma^2 \mu^4 {\rm e}^{\frac{q^2-Q^2}{\mu^2}} = {\rm const} = 0 \,,
  \label{exp2}
\end{equation}
so that its implicit solution
\begin{equation}
  u=\frac{1}{\gamma P}\sqrt{\frac{\pi}{2}}\,{\rm e}^{\frac{Q^2}{2\mu^2}}\left[\erf\left(\frac{q}{\mu\sqrt2}\right)-\erf\left(\frac{Q}{\mu\sqrt2}\right)\right]
  \label{explicitsol2}
\end{equation}
is expressed in terms of the Gauss error function $\erf(x)$, defined as
\begin{equation}
\erf(x)=\frac{2}{\sqrt{\pi}}\int\limits_0^x dt\, {\rm e}^{-t^2} \,, \quad \erf(-x) = - \erf(x) \,.
\label{Gauss}
\end{equation}
When $|q|=\infty$, the first Gauss error function in (\ref{explicitsol2}) takes finite value; this means that there exists a finite value $u_{\infty}$, for which $|q(u_{\infty})|=\infty$.
For instance, when $Q$ is positive, $q(u_{\infty})= -\infty$, and $u_{\infty}$ can be found as follows:
\begin{equation}
  u_{\infty}=\frac{1}{\gamma |\mu|}\sqrt{\frac{\pi}{2}}\,{\rm e}^{\frac{Q^2}{2\mu^2}}\left[1+\erf\left(\frac{|Q|}{|\mu|\sqrt2}\right)\right] >0 \,.
  \label{explic}
\end{equation}
The radial function $r$ (\ref{r}) also can be presented in terms of Gauss error functions:
\begin{align}
  r={\rm e}^{\beta-\rho}&=\sqrt{\frac{2}{\pi}}\gamma P {\rm e}^{-\frac{q^2}{2\mu^2}}\nonumber\\
  {}&{}\times\left[\erf\left(\frac{q}{\mu\sqrt2}\right)-\erf\left(\frac{Q}{\mu\sqrt2}\right)\right]^{-1}. \label{explicitsolr}
\end{align}
According to this formula, $r(u_\infty)=0$, thus, we obtain that  $E(u_\infty)= \infty$ and $\phi(u_\infty)= \infty$.
In other words, we deal with central singularity at $u=u_\infty$.
On the other hand, $q=0$, when $u=u_{0}$, where
\begin{equation}
u_0=\frac{1}{\gamma |\mu|}\sqrt{\frac{\pi}{2}}\,{\rm e}^{\frac{Q^2}{2\mu^2}}\erf\left(\frac{|Q|}{|\mu|\sqrt2}\right) \,,
\label{Gauss2}
\end{equation}
which is valid for arbitrary signs of $Q$ and $\mu$. Similarly, we obtain
\begin{equation}
  r(u_0)= \sqrt{\frac{2}{\pi}}\gamma |\mu|  \left[\erf\left(\frac{|Q|}{|\mu|\sqrt2}\right)\right]^{-1}>0 . \label{2explicitsolr}
\end{equation}
Thus, the electric field takes zero value, when $u=u_0$ and $r=r(u_0)$. Since $E(u=u_0)=E(u=0)=0$, the function $E(u)$ reaches extremum at the finite value of the variable $u$ (the type of extremum, minimum or maximum, is predetermined by the sign of the electric charge $Q$). Typical plots of $E(r)$ and $\phi(r)$ are presented in Fig.~\ref{Fig1}.

\subsection{Axion electrostatics in the background field of the magnetic monopole}\label{backgroundsolution}

\subsubsection{Preamble: The regular solution to the equation of axion electrostatics in the flat spacetime}

In order to simplify further interpretation of solutions, let us assume, first, that the background spacetime is flat, i.e., $\beta=0$ and $\rho=\ln u$. Then the last equation in (\ref{keyrep1}) reduces to the form
\begin{equation}
  q''_{uu}-\gamma^2\mu^2q=0,
\end{equation}
and we can obtain the solution discovered by Campbell, Kaloper, and Olive \cite{Campbell}, which is regular at the center $u=\infty$ and satisfies the condition $q(u=0)=Q$
\begin{equation}
  q(u)=Q {\rm e}^{-\gamma|\mu| u}. \label{CKO}
\end{equation}
Another boundary condition $q'(0)=\gamma \mu P$ gives the constraint on the axion charge
\begin{equation}
P=-Q \sgn \mu,\label{PQ-1}
\end{equation}
for which this regular solution exists.

\subsubsection{Exact solution to the equation of axion electrostatics}

We assume now that the background gravitational field is formed by the magnetic monopole without horizons and with the naked singularity at the center. In fact, the background metric relates to the Reissner-Nordstr\"om solution with a magnetic charge.
This means that $C= -\nu^2 <0$, and
\begin{equation}
{\rm e}^\rho=\Pi(C,u)=\frac{\sin \nu u}{\nu}, \quad \nu=\sqrt{R_\mu^2-M^2} \,,
\label{n1}
\end{equation}
\begin{equation}
 {\rm e}^\beta= R_{\mu}\,\frac{\sin \nu(u+u_*)}{\nu}, \quad \sin \nu u_*=\frac{\nu}{R_\mu} \,.
\label{n2}
\end{equation}
\begin{equation}
  r={\rm e}^{\beta-\rho}=  R_\mu\, \frac{\sin \nu (u+u_*)}{\sin \nu u},
\label{n3}
\end{equation}
where $R_\mu=\sqrt{\frac{\kappa \mu^2}{2}}$ is the Reissner-Nordstr\"om radius.
When $u+u_* \to \frac{\pi}{\nu}$, we obtain that ${\rm e}^\beta \to 0$ and $r \to 0$.

In this spacetime background the function $q(u)$, which determines the electric field induced by the axion-photon coupling, satisfies the equation
\begin{equation}
 \sin^2 \nu (u+u_*) \ q_{uu} = \frac{\gamma^2\mu^2 \nu^2}{(\nu^2+M^2)} \ q .
\label{n4}
\end{equation}
The replacement $z=i\cot \nu(u+u_*)$ transforms this equation into the Legendre  equation
\begin{equation}
 \frac{d}{dz}\left[(1-z^2)\frac{dq}{dz} \right] + \alpha (\alpha +1) q =0.
\label{n5}
\end{equation}
where the parameter $\alpha$ is introduced as follows:
\begin{gather}
\alpha (\alpha +1)= \frac{\gamma^2\mu^2}{(\nu^2+M^2)}=\frac{2\gamma^2}{\kappa} , \nonumber\\
\alpha=\frac12 \left(\pm \sqrt{1+\frac{8\gamma^2}{\kappa}}-1\right)\,.
\label{n6}
\end{gather}
The variable $z$ is complex; the quantity $|z|$ takes the value $|z|=\frac{M}{\nu}$ at $u=0$, and becomes infinite $|z|=\infty$, when $u=\frac{\pi}{\nu}-u_{*}$. We search for the solution $q(z)$, which is regular for the interval $\frac{M}{\nu} < |z| < \infty$, and we require especially, that the solution is regular at $u=\frac{\pi}{\nu}-u_*$. As usual, $q(z)$, the solution of the Legendre equation (\ref{n5}), is the linear combination of ${\sf P}_\alpha(z)$ and ${\sf Q}_\alpha(z)$, the Legendre functions of the first and second kinds, respectively (see, e.g.,  \cite{Erd} for details). Keeping in mind the analytic properties of the Legendre functions, we can write the regular solution for $q(u)$ satisfying the condition $q(0)=Q$ in the following form:
\begin{gather}
  \frac{q(u)}{Q} = \frac{{\sf Q}_\alpha(z(u))+\pi\, i\, {\sf P}_\alpha(z(u))\Theta(\cot \nu (u+u_*))}
  {{\sf Q}_\alpha(i\cot \nu u_*)+\pi\, i\, {\sf P}_\alpha(i\cot \nu u_*)} ,\nonumber\\
  z(u)=i\cot\nu(u+u_*).\label{qq1}
\end{gather}
Here $\Theta(\cot \nu (u+u_*))$ is the Heaviside function; as was shown in \cite{ExtendedAE5} such a structure guarantees the regularity of the solution on the real axis of the complex plane $z$.
Using (\ref{qq1}) and  (\ref{n3}) we can present the electric field $E(r)$ as a function of $r$ as follows:
\begin{gather}
E(r)=\frac{Q}{r^2}\cdot \frac{{\sf Q}_\alpha\!\left(\frac{iM}{\nu}\Xi(r)\right)+\pi\, i\, {\sf P}_\alpha\!\left(\frac{iM}{\nu}\Xi(r)\right)\Theta(\Xi(r))}
{{\sf Q}_\alpha\!\left(\frac{iM}{\nu}\right)+\pi\, i\, {\sf P}_\alpha\!\left(\frac{iM}{\nu}\right)} \,,\nonumber\\
\Xi(r)=1-\frac{R_\mu^2}{rM}.\label{Emm3}
\end{gather}


\subsubsection{Integral representation of the solution}

For the analysis of regularity of the electric field one can use also the convenient integral representations of the Legendre functions (see \cite{Erd})
\begin{gather}
{\sf Q}_\alpha(i\cot x)+\pi\, i\, {\sf P}_\alpha(i\cot x)\Theta(\cot x)\nonumber\\
{}=\frac{{\rm e}^{\frac{i\pi(\alpha+1)}{2}}}{(\sin x)^\alpha}\int\limits_x^\pi d\xi \left(\cos x- \cos \xi\right)^\alpha\nonumber\\
{}=\frac{{\rm e}^{\frac{i\pi(\alpha+1)}{2}}}{(\sin x)^\alpha}\int\limits_{-\cos x}^1 dz \frac{\left(z+\cos x\right)^\alpha}{\sqrt{1-z^2}},\label{int1}
\end{gather}
which yields, in particular,
\begin{gather}
{\sf Q}_\alpha(i\cot \nu u_*)+\pi\, i\, {\sf P}_\alpha(i\cot \nu u_*)\nonumber\\
{}={\rm e}^{\frac{i\pi(\alpha+1)}{2}}\left(1+\frac{M^2}{\nu^2}\right)^{\frac{\alpha}{2}}\!\int\limits_{-\frac{M}{\sqrt{\nu^2+M^2}}}^1 \! dz \frac{\left(z+\frac{M}{\sqrt{\nu^2+M^2}} \right)^\alpha}{\sqrt{1-z^2}} \,.
\label{int2}
\end{gather}


Using these representations, one can show that
\begin{equation}
\frac{q}{Q}=N^{\alpha/2}\cdot
\frac{Z_\alpha\left(\left[\frac{M}{R_\mu}-\frac{R_\mu}{r}\right]N^{-1/2}\right)}{Z_\alpha(\frac{M}{R_\mu})},
\label{mm3}
\end{equation}
where the function $Z_\alpha(\xi)$ is defined as follows
\begin{equation}
Z_\alpha(\xi)=\int\limits_{-\xi}^1 \frac{dt \left(t+\xi \right)^\alpha}{\sqrt{1-t^2}} \,,
\label{mm2}
\end{equation}
and $N$ is the standard Reissner-Nordstr\"om metric coefficient
\begin{equation}
 N=1-\frac{2M}{r}+\frac{R_\mu^2}{r^2}.
\label{mm4}
\end{equation}
The function $Z_\alpha(\xi)$ satisfies the following relations:
\begin{gather}
Z_{\alpha}(0)=\frac{\sqrt{\pi}}{2}\cdot\frac{\Gamma(\frac{\alpha+1}{2})}{\Gamma(\frac{\alpha+2}{2})},\\ Z_{\alpha}(1)=\frac{2^{\alpha}\sqrt{\pi}\,\Gamma(\alpha+\frac{1}{2})}{\Gamma(\alpha+1)},\label{mm72}\\
Z_{\alpha}(-1+\xi)\sim \sqrt{\frac{\pi}{2}}\cdot\frac{\Gamma(\alpha+1)}{\Gamma(\alpha+\frac{3}{2})} \, \xi^{\alpha+\frac{1}{2}},\label{mm71}\\
\left.Z_{\alpha}(\xi)\right|_{\alpha\to\infty}\sim \sqrt{\frac{\pi}{2\alpha}}\,(1+\xi)^{\alpha+\frac12}.\label{mm73}
\end{gather}
Using the formula (\ref{mm71}), one can obtain that
\begin{equation}
\left. \frac{q}{Q}\right|_{r\to 0}\sim \sqrt{\pi}\cdot\frac{\Gamma(\alpha+1)}{\Gamma(\alpha+\frac{3}{2})}
\frac{\left(1-\frac{M^2}{R_\mu^2}\right)^{\alpha+1/2}}{2^{\alpha+1} Z_\alpha(\frac{M}{R_\mu})}\left(\frac{r}{R_\mu}\right)^{\alpha+1} \,.
\label{mm1}
\end{equation}
\begin{figure}[t]
\centerline{\includegraphics[height=7.5cm]{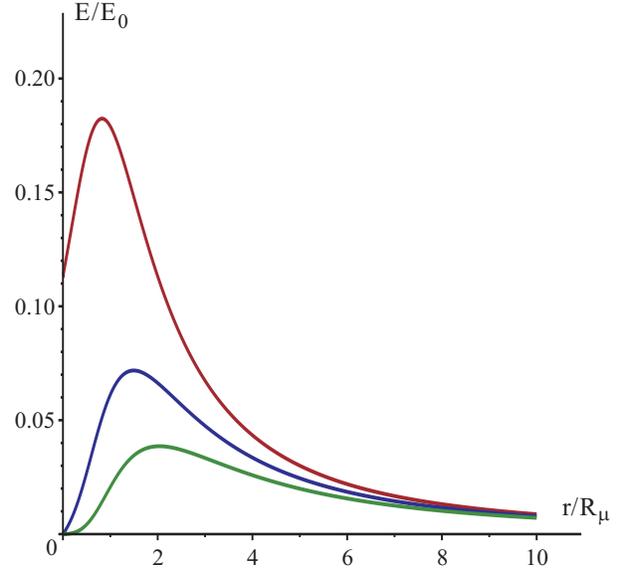}}\caption{Plots of the function $E(r)/E_0$, where $E_0=\frac{Q}{R_\mu^2}$, for the cases, when the electric field is regular at the center. The upper curve relates to the critical value of the parameter $\alpha$, $\alpha=1$; the electric field at the center is finite but nonvanishing for $\alpha=1$. Two other plots relate to the values  $\alpha=2$ and $\alpha=3$ (bottom curve); the electric field takes zero value at the center in these cases. For presented plots the mass value is chosen to be equal to $M=0.5R_\mu$.}\label{fig2}
\end{figure}
The electric field $E(r)=\frac{q}{r^2}$ is regular at the center $r=0$, when $\alpha \geq 1$. The value $E(0)$ is finite, when $\alpha= 1$, and $E(0)=0$, when  $\alpha>1$.
The second invariant of the electromagnetic field $I_{(2)}\equiv \frac{1}{4}F_{mn}\Fst^{mn}$ is regular at the center, when $\alpha \geq 3$. Indeed,
\begin{equation}
 \frac14 F_{mn}\Fst^{mn} = \frac{1}{\sqrt{-g}} E^{ut \theta \varphi} F_{ut} F_{\theta \varphi} = \mu q {\rm e}^{4\rho-4\beta}=\frac{\mu q}{r^4},
\label{mm4}
\end{equation}
thus, at $r\to 0$ the invariant $\frac14 F_{mn}\Fst^{mn} \propto r^{\alpha-3}$. In Fig.~\ref{fig2} we present typical plots of the function $E(r)$ for three values of the parameter $\alpha$.

\subsubsection{Behavior of the axion field}

When the function $q(u)$ is found, the axion field $\phi=\phi_0 + \frac{q-Q}{\gamma\mu}$ can be easily reconstructed. In particular, one can see that the axion field is regular at the center, when the function $q(u)$ takes finite value at $r=0$. We focus now on the following detail: when $u \to 0$, the quantity $\frac{q-Q}{\gamma\mu}$ tends to $\frac{u q^{\prime}(0)}{\gamma\mu}$, so, in fact, we have to analyze the value of the quantity $\frac{P}{Q}$. This ratio can be calculated as
\begin{align}
  \frac{P}{Q}&=-\sqrt{\frac{\alpha+1}{\alpha}}\sgn\mu \nonumber\\
  {}&\times\left(\frac{\nu}{iR_\mu}\cdot\frac{{\sf Q}_{\alpha+1}\!\left(\frac{iM}{\nu}\right)+\pi\, i\, {\sf P}_{\alpha+1}\!\left(\frac{iM}{\nu}\right)}{{\sf Q}_\alpha\!\left(\frac{iM}{\nu}\right)+\pi\, i\, {\sf P}_\alpha\!\left(\frac{iM}{\nu}\right)}-\frac{M}{R_\mu}\right).\label{PQration1}
\end{align}
Integral representation (\ref{int1}) of this quantity gives
\begin{equation}
\frac{P}{Q}=-\sqrt{\frac{\alpha+1}{\alpha}}\sgn\mu \cdot\left( \frac{Z_{\alpha+1}(\frac{M}{R_\mu})}{Z_\alpha(\frac{M}{R_\mu})}-\frac{M}{R_\mu}\right).
\end{equation}
For two limiting cases, $M\to0$ and $M\to R_\mu$, this expression takes the form
\begin{equation}
  \left.\frac{P}{Q}\right|_{M\to 0}=-\frac{2\sgn\mu}{\sqrt{\alpha(\alpha+1)}}
  \cdot\left[\frac{\Gamma\!\left(\frac{\alpha}{2}+1\right)}{\Gamma\!\left(\frac{\alpha+1}{2}\right)}\right]^2,
\end{equation}
\begin{equation}
  \left.\frac{P}{Q}\right|_{M\to R_\mu}=-\sqrt{\frac{\alpha}{\alpha+1}}\sgn\mu.
\end{equation}
 Using the formula (\ref{mm73}), one can demonstrate the following detail: if $\alpha\to\infty$, i.e., if the axion-photon coupling constant $\gamma$ is much greater than $\sqrt{\kappa}$, the ratio $P/Q$ tends to a constant for any values of $M\in[0,R_\mu)$:
\begin{equation}
  \left.\frac{P}{Q}\right|_{\alpha\to\infty}\to -\sgn\mu.
\end{equation}
Thus, in this limit, $\alpha\to\infty$, we obtain the result coinciding with the flat spacetime case  (see~(\ref{PQ-1})).

\subsubsection{Limiting case $M\to R_\mu$}

Let us consider the extremal Reissner-Nordstr\"om case with $M\to R_\mu$. For this limit, we have
\begin{equation}\label{RNlimit}
  \nu=\sqrt{R^2_\mu-M^2}=0,\quad u_*=\frac{1}{R_\mu}, \quad u=\frac{1}{r-R_\mu}.\nonumber
\end{equation}
The metric function $\beta$ takes the form
\begin{equation}\label{RNlimit1}
  \beta=-\ln\left(1-\frac{R_\mu}{r}\right),
\end{equation}
while the electric field function $q$ according the formula (\ref{mm3}) can be written as follows
\begin{equation}\label{RNlimit1}
  q=Q\left(1-\frac{R_\mu}{r}\right)^\alpha,
\end{equation}
or, excluding the variable $r$,
\begin{equation}\label{RNlimit2}
  \beta =-\frac{1}{\alpha}\ln\frac{q}{Q}.
\end{equation}
In contrast to the case $M<R_\mu$, the function $q$ vanishes at the double horizon $r=R_\mu\neq0$ and ${\rm e}^\beta|_{q=0}\to\infty$.

\subsection{Qualitative and numerical studies of the regular solutions}\label{numsolution}

When the spacetime background is not fixed, i.e., the model is self-consistent, we have to solve the general system of the key equations (\ref{keyrep1}) and (\ref{vv}). In contrast to the explicit example demonstrated in Subsection~\ref{exactS}, regular solutions to this system can be presented in a numerical form only.

In this subsection we will study solutions with the electric field, regular at the center, when the function $q$ vanishes at $r=0$. The metric function $\beta$ has to tend to $-\infty$, because the naked singularity associated with the magnetic monopole cannot be removed.
Using Eq.~(\ref{vv}), we obtain that $\beta$ behaves as
\begin{equation}
\left.\beta\right|_{q\to0}\sim \frac{1}{\alpha+1}\ln \frac{q}{Q},\label{asymBN}
\end{equation}
where the parameter $\alpha>-1$ has to satisfy the condition
\begin{gather}
\alpha(\alpha+1)=\frac{2\gamma^2}{\kappa},\nonumber \\
\alpha=\frac12\left(\sqrt{1+\frac{8\gamma^2}{\kappa}}-1\right).\nonumber
\end{gather}
Obviously, this relation does not differ from the corresponding expression (\ref{n6}) for the background solution.
The formula (\ref{asymBN}) relates to the following asymptotic behavior of the functions $q(u)$ and $\beta(u)$:
\begin{equation}
\beta(u)\propto \ln (u_0-u),\qquad q(u)\propto (u_0-u)^{\alpha+1},\label{asymBqN}
\end{equation}
where the value $u_0$ corresponds to the value  $r=0$ at the center. For instance, for the background solution considered above $u_0=\frac{\pi}{\nu}-u_*$.
When ${\rm e}^{\rho}$ does not vanish at $u=u_0$, the standard radial coordinate $r={\rm e}^{\beta-\rho}$ behaves as follows:
$$r\propto u_0-u,$$
and we have
\begin{gather}
q\propto r^{\alpha+1}, \quad \beta\propto \ln r.\label{asymBqR}
\end{gather}
The first formula coincides qualitatively with the corresponding expression for the background solution (see (\ref{mm1}), and the electric field $E(r) = \frac{q}{r^2}$ vanishes at the center when $\alpha>1$ as well.

\begin{figure*}
\begin{tabular}{rl}
{\includegraphics[height=6.0cm]{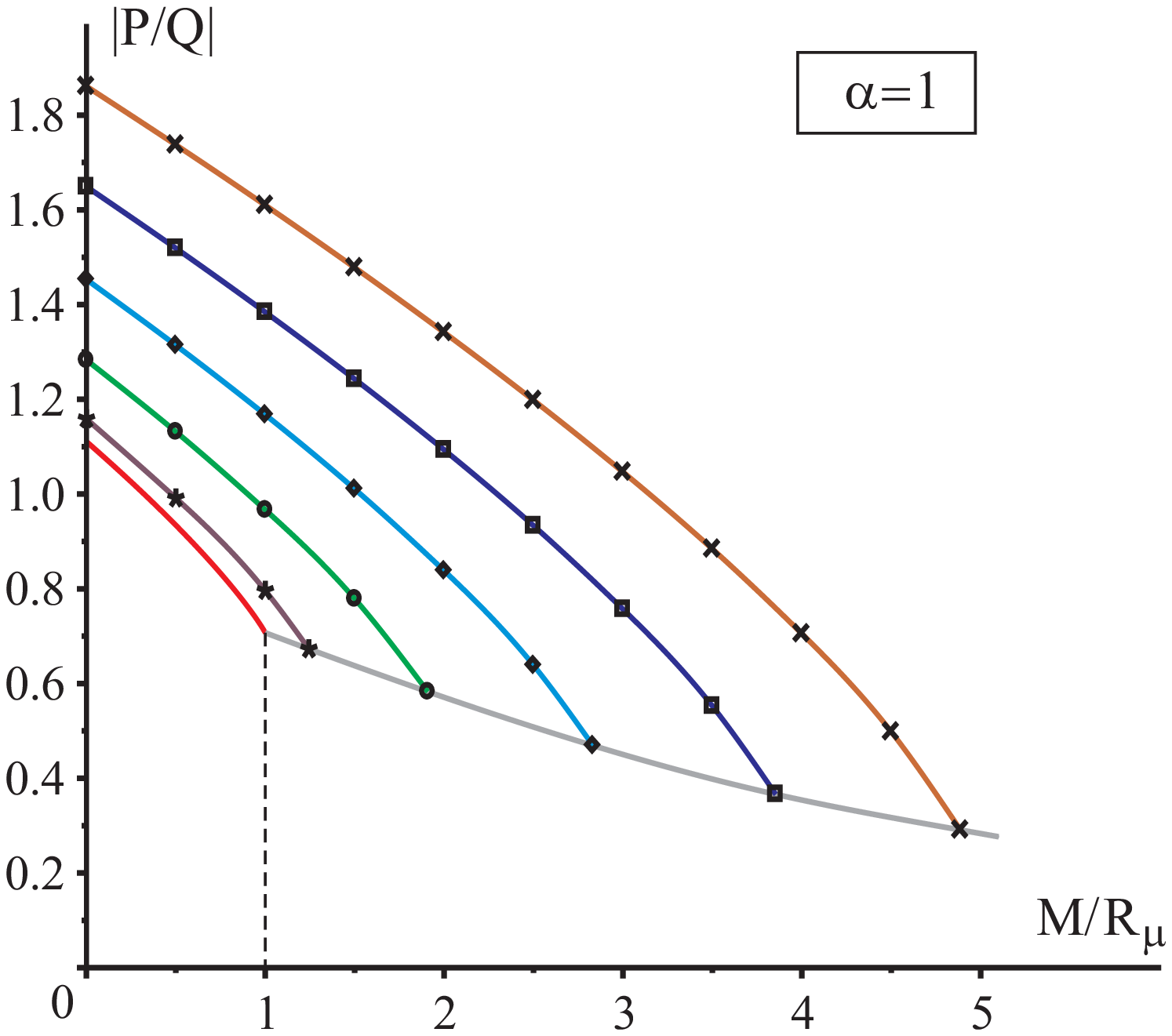}}\hspace{0.5cm} & \hspace{0.5cm}{\includegraphics[height=6.0cm]{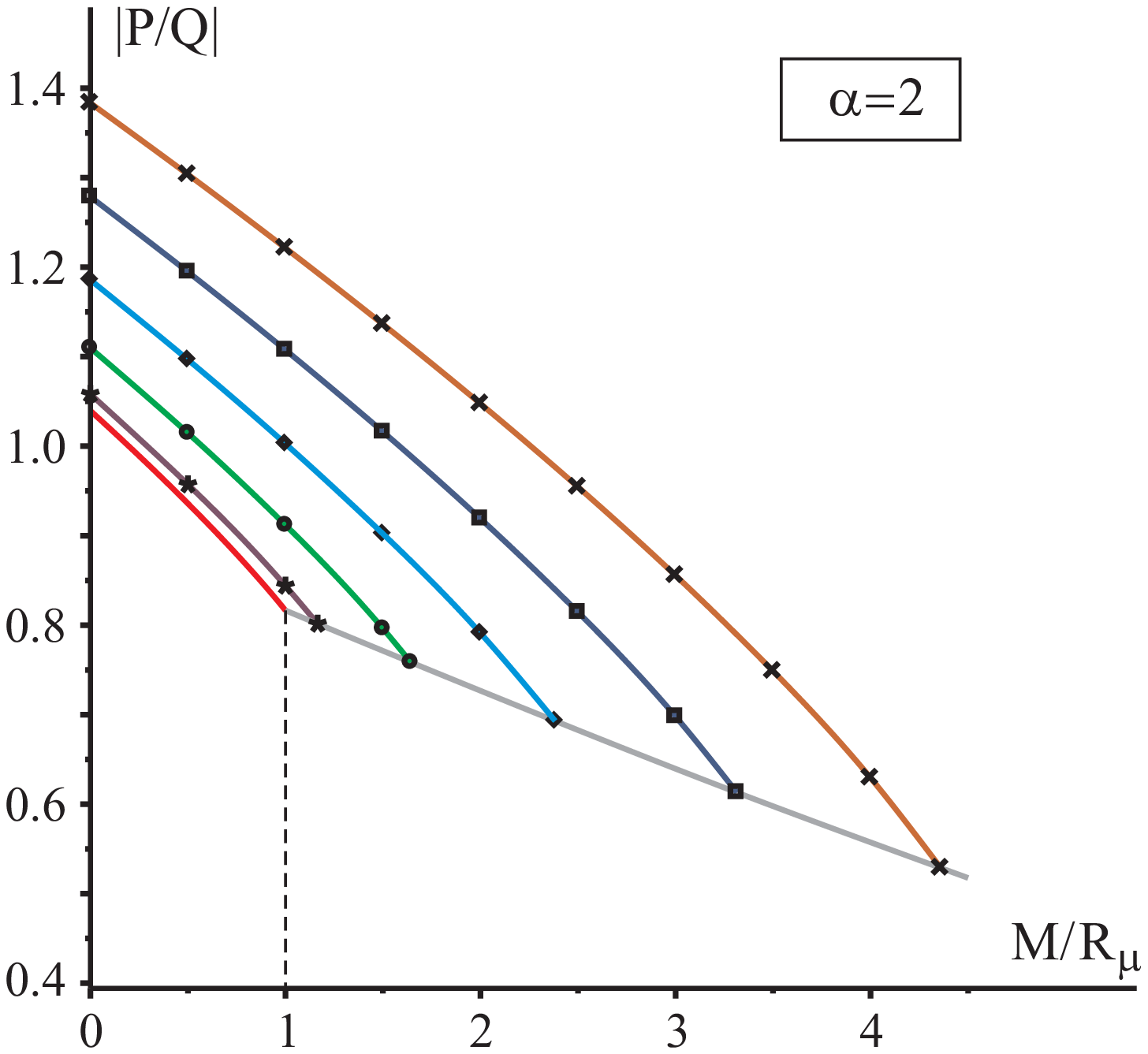}} \cr
{\vspace{0.5cm}} & {} \cr
\end{tabular}
\begin{tabular}{rl}
{\includegraphics[height=6.0cm]{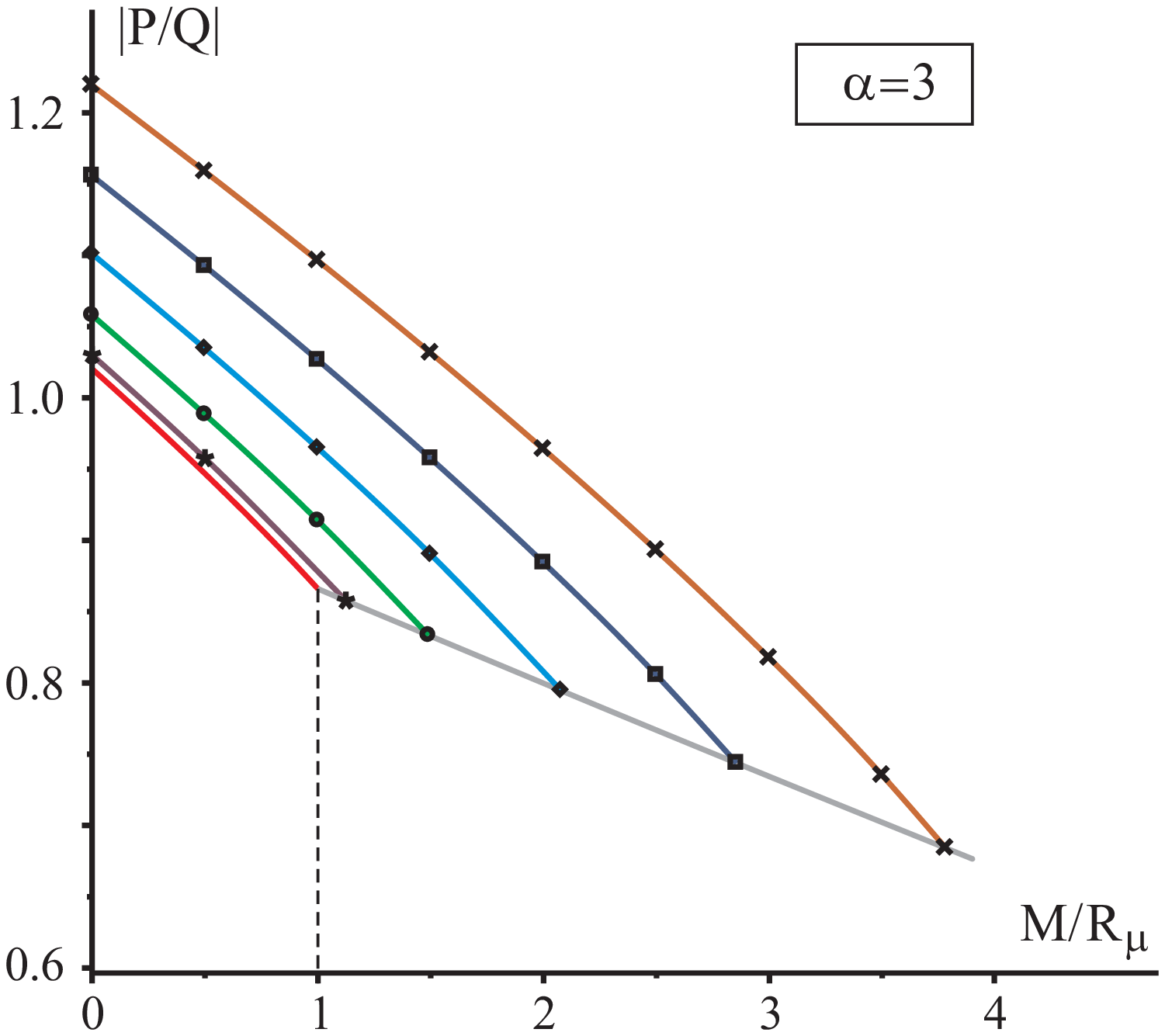}}\hspace{0.5cm} & {\includegraphics[height=6.0cm]{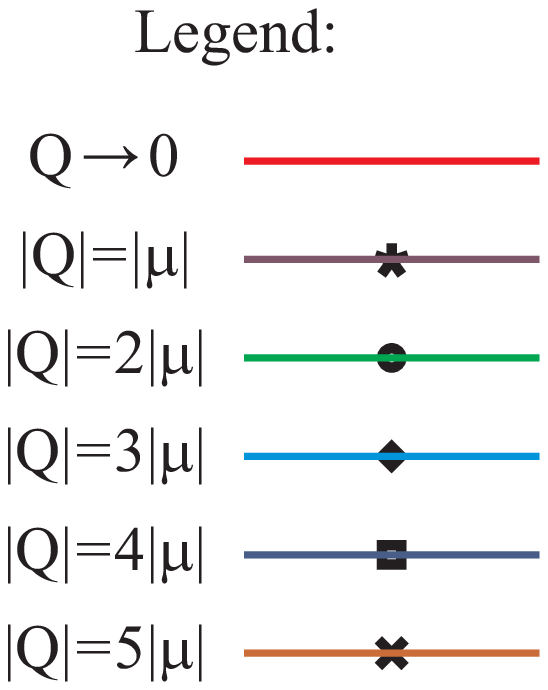}}
\end{tabular}
\caption{Dependence between the ratio $P/Q$ and the dimensionless mass $M/R_\mu$, where $R_\mu=\kappa \mu^2/2$, for the solution with the regular electric field. Panels relate to specific values of the coupling parameter $\alpha=1,\dots,3$. The first (left) line corresponds to the limiting case $Q\ll\mu$. Other lines correspond to $|Q|=n|\mu|$, where $n=1,\dots ,5$. The gray line defines the relationship between the mass and the charges $P$ and $Q$ for the spacetime metric with double (extremal) horizon.}\label{pics}
\end{figure*}

\begin{figure*}
\begin{tabular}{rl}
{\includegraphics[height=6.0cm]{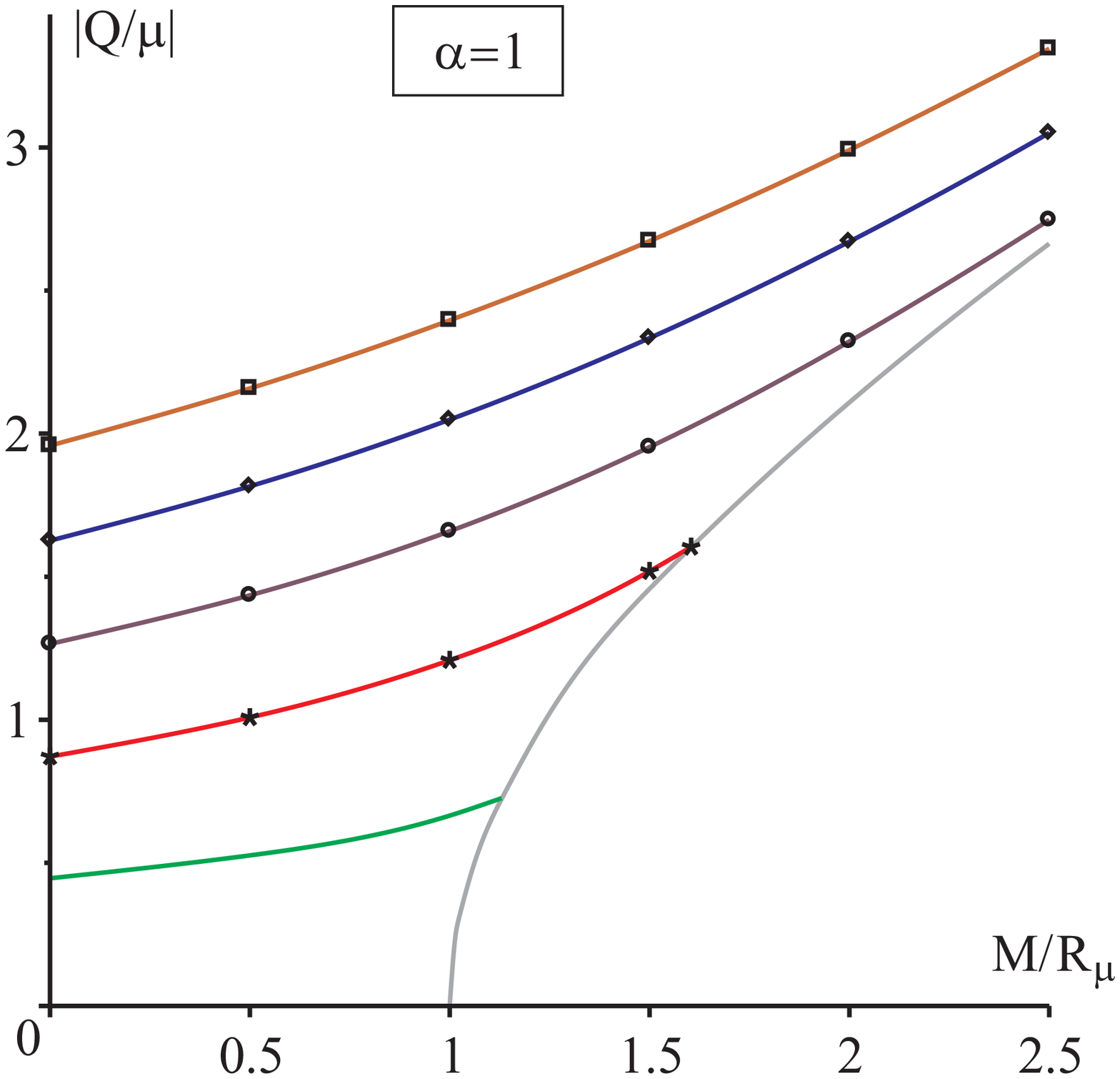}}\hspace{0.5cm} & \hspace{0.5cm}{\includegraphics[height=6.0cm]{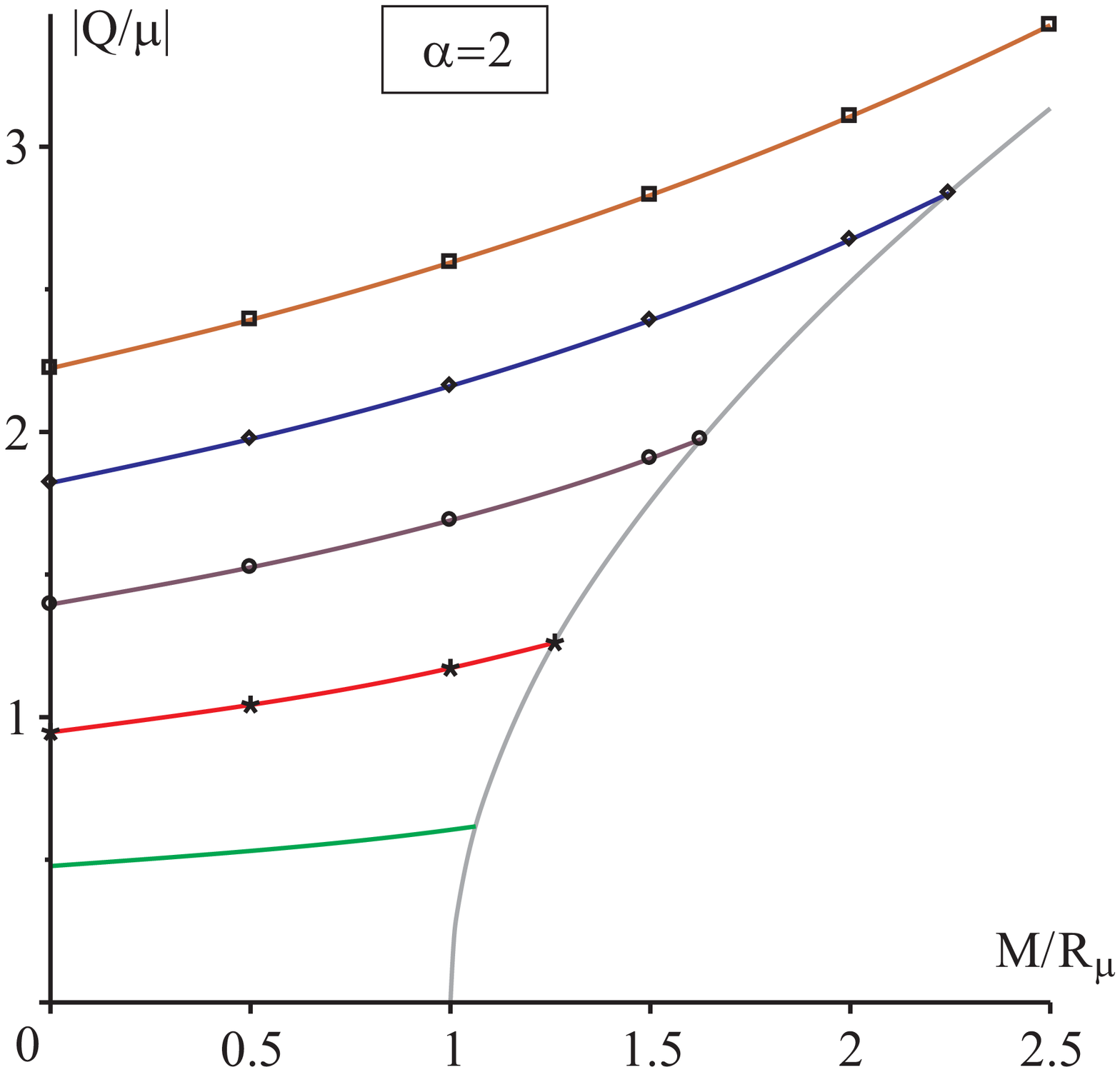}} \cr
{\vspace{0.5cm}} & {} \cr
\end{tabular}
\begin{tabular}{rl}
{\includegraphics[height=6.0cm]{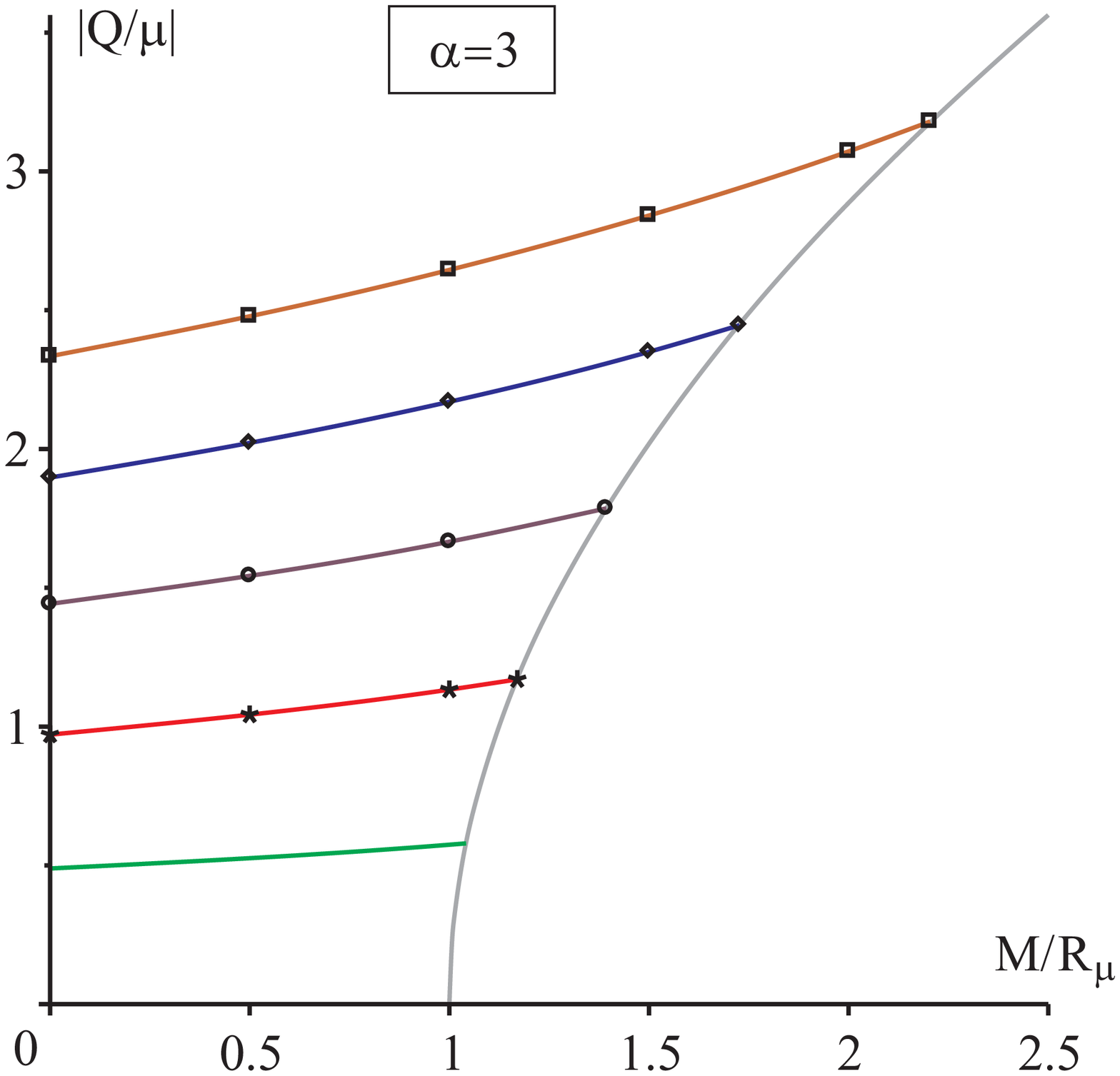}}\hspace{0.5cm} & {\includegraphics[height=6.0cm]{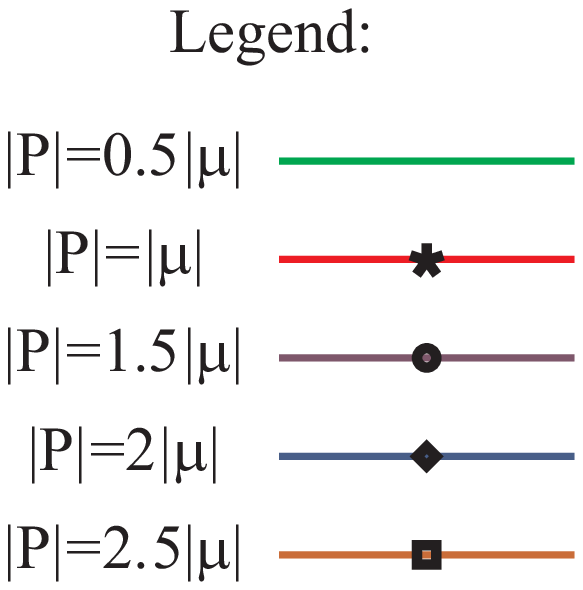}}
\end{tabular}
\caption{Dependence between the ratio $Q/\mu$ and the dimensionless mass $M/R_\mu$, where $R_\mu=\kappa \mu^2/2$, for the solution with the regular electric field. Panels relate to specific values of the coupling parameter $\alpha=1,\dots,3$. The range $M/R_\mu \in [0,1)$ on the horizontal axis corresponds to the limiting case $P\ll\mu$. Lines on the panel correspond to $|P|=0.5n|\mu|$, where $n=1,\dots ,5$. The gray line defines the mass-charge relation for the spacetime metric with double (extremal) horizon.}\label{qms}
\end{figure*}

When $u=0$ the boundary conditions (\ref{k1}) give
\begin{equation}
  \beta|_{q=Q}=0,\qquad \left.\beta'_q\right|_{q=Q}=\frac{M}{\gamma\mu P}.\label{kk1}
\end{equation}
If we fix the electric and magnetic charges, $Q$ and $\mu$, the coupling constant $\gamma$, and the mass $M$, desired solution to Eq.~(\ref{vv}) with conditions (\ref{asymBN}) and (\ref{kk1}) exists only for a specific value of the axion charge $P$, and the inequality $\mu P/Q<0$ has to be valid. This latter constraint arises from the second equation of (\ref{keyrep1}), because
$$\frac{q(0)}{Q}=1,\quad \frac{q(u_0)}{Q}=0,\quad \frac{q''_{uu}}{Q}=\frac{\gamma^2\mu^2q\,{\rm e}^{-2\beta}}{Q}>0.$$
To illustrate dependence between $Q$, $\mu$, $P$, and $\alpha$ (or $\gamma$), we present Figs.~\ref{pics} and~\ref{qms}. Each figure consists of three panels, which correspond to specific values of the coupling parameter $\alpha$, namely, for $\alpha=1$, 2, and 3, respectively.

On the other hand, if $C=0$ Eq.~(\ref{vv}) admits a solution, which at $q=0$ behaves as follows (cf.~(\ref{RNlimit2}))
\begin{equation}
  \beta\sim -\frac{1}{\alpha}\ln \frac{q}{Q}.
\end{equation}
As it was mentioned above, such a solution corresponds to the metric, which possesses a double horizon. Curves describing in Figs.~\ref{pics} and \ref{qms} the relationship between the charges, $P$, $Q$, and $\mu$, and the mass $M$ for this {\it limiting} configuration are drawn using gray color.

Fig.~\ref{pics} depicts dependence between the ratio $P/Q$ and the mass $M$ for fixed values of the electric charge $Q$ and the coupling parameter $\alpha$. The first (left) curve corresponds to the limiting case $Q\ll\mu$ described in Subsection~\ref{backgroundsolution}. Other curves correspond to $|Q|=n|\mu|$, where $n=1,\dots ,5$.

Fig.~\ref{qms} illustrates dependence between the ratio $Q\mu$ and the mass $M$ for fixed values of the axion scalar charge $P$ and the coupling parameter $\alpha$. The range $M/R_\mu \in [0,1)$ on the horizontal axis corresponds again to the limiting case $P\ll\mu$ described in Subsection~\ref{backgroundsolution}. Color curves correspond to $|P|=0.5n|\mu|$, where $n=1,\dots ,5$. The gray line defines the mass-charge relation for the spacetime metric with double (extremal) horizon. If the gravitational interaction is much weaker than the axion-photon coupling, i.e. when $\gamma^2\gg\kappa$, or, equivalently, $\alpha\gg 1$, the color curves become horizontal, $|P|\to |Q|$ (see (\ref{PQ-1})). The solution of Campbell, Kaloper, and Olive (\ref{CKO}) can be considered as a non-gravitational limit.

\section{Conclusions}\label{conclusion}

In the present paper we realize Wilczek's idea about a magnetic monopole surrounded by an axion-induced radial
electric field in the framework of the Einstein-Maxwell model with the massless axion field. Since this electric field is created by interaction between the magnetic field of the monopole and the axion field and is not related to any real electric charge, the electric field has to be regular in the center in the strict sense, i.e., $E(0)=0$. In this sense, our solution is a generalization of the result of Campbell, Kaloper, and Olive \cite{Campbell}, taking into account the gravitational field of the monopole.

In Subsection~\ref{backgroundsolution} we present the four-parameter family of solutions (see Eqs.~(\ref{Emm3}) and (\ref{mm3})) in the framework of the axion electrodynamics on the background of the magnetic monopole gravitational field with the metric of the Reissner-Nordstr\"om type. The fifth parameter, the axion field charge $P$, is determined by other parameters, namely, the electric and magnetic charges $Q$ and $\mu$, the mass $M$, and the coupling parameter $\alpha$ (see Eq.~(\ref{PQration1})). Besides this relation, the parameters are bounded by two inequalities, which correspond to requirements of absence of horizons ($M^2<\kappa \mu^2/2$) and regularity at the origin ($\alpha>1$). In addition, when $\alpha\geq 3$ the invariant scalar $F_{ik}\Fst^{ik}$ appears to be regular in the center too.

In Subsection~\ref{numsolution}, using numerical methods, we solve  the total system of equations attributed to the Einstein-Maxwell-axion model, in which the gravitational field is self-consistent, not the background one. We demonstrate that the behavior of the solutions to the self-consistent system qualitatively coincides with the background solution, and this background solution can be extracted from the general solution as an asymptotic case with $Q,\ P\ll\mu$.

\appendix

\acknowledgments
The work was supported by Russian Science Foundation (Project No. 16-12-10401), and, partially, by the Program of Competitive Growth of Kazan Federal University.

\end{document}